\documentclass[letterpaper,twocolumn,10pt]{article}
\usepackage{usenix-2020-09}
\usepackage[toc,page]{appendix}
\usepackage{filecontents}
\usepackage{tikz}
\usepackage{amsmath}
\usepackage{multirow}
\usepackage{graphicx}
\usepackage{subcaption}
\usepackage{listings}
\usepackage{xcolor}
\usepackage{multicol}
\usepackage{tcolorbox}
\usepackage{array}
\tcbuselibrary{breakable}

\usepackage{csquotes}
\usepackage{adjustbox}
\usepackage{array}
\newcommand{\PreserveBackslash}[1]{\let\temp=\\#1\let\\=\temp}
\newcolumntype{C}[1]{>{\PreserveBackslash\centering}p{#1}}
\newcolumntype{R}[1]{>{\PreserveBackslash\raggedleft}p{#1}}
\newcolumntype{L}[1]{>{\PreserveBackslash\raggedright}p{#1}}
\usepackage{multirow,bigdelim}
\usepackage{booktabs}
\usepackage{tabularx}
\usepackage{makecell}
\usepackage{enumitem}

\lstdefinelanguage{yaml}{
  keywords={true,false,null,y,n},
  keywordstyle=\color{blue}\bfseries,
  basicstyle=\ttfamily\small,
  comment=[l]{\#},
  morecomment=[s]{/*}{*/},
  commentstyle=\color{gray}\ttfamily,
  stringstyle=\color{red}\ttfamily,
  morestring=[b]',
  morestring=[b]",
  sensitive=false
}

\usepackage{filecontents}
\usepackage{authblk}

\newcommand{\name}{TIPS}
\newcommand{\stm}{SecEncoder}

\newcommand{\baseline}{\name$_{General}$}
\newcommand{\focused}{\name$_{Focused}$}

\begin{document}

\date{}

\title{\Large \emph{\name{}:} \bf Threat Actor Informed Prioritization of Applications using SecEncoder}

\author{Muhammed Fatih Bulut}
\author{Acar Tamersoy}
\author{Naveed Ahmad}
\author{Yingqi Liu}
\author{Lloyd Greenwald}
\affil{Microsoft Security AI Research \authorcr {\tt \{mbulut, acartamersoy, navahm, yingqiliu, lgreenwald\}@microsoft.com}}

\maketitle

\begin{abstract}
This paper introduces \name{}: Threat Actor Informed Prioritization using \emph{SecEncoder}, a specialized language model for security. \name{} combines the strengths of both encoder and decoder language models to detect and prioritize compromised applications. By integrating threat actor intelligence, \name{} enhances the accuracy and relevance of its detections. Extensive experiments with a real-world benchmark dataset of applications demonstrate \name{}'s high efficacy, achieving an F-1 score of 0.90 in identifying malicious applications. Additionally, in real-world scenarios, \name{} significantly reduces the backlog of investigations for security analysts by 87\%, thereby streamlining the threat response process and improving overall security posture.
\end{abstract}
\section{Introduction}
\label{sec:intro}

Since the introduction of transformers \cite{vaswani2017attention}, foundational models have become pivotal, significantly influencing various facets of our lives. Prominent examples include encoder-only models like BERT and DeBERTa \cite{devlin2019bert, he2021deberta}, as well as decoder-only models such as GPT, Gemini, and Llama \cite{openai2023gpt4, 2024gemini15, touvron2023llama}. These models leverage extensive datasets and possess the capability to reason across diverse tasks, transforming fields such as natural language processing, computer vision, and more.

Security has always been at the forefront of adopting machine learning-based solutions, addressing tasks such as anomaly detection, phishing detection, intrusion detection, and various other critical functions. With the general availability of large language models (LLMs), we have already begun to see their integration into everyday security tasks, including incident enrichment, summarization, vulnerability detection, and log analysis \cite{stm_paper_2024, charalambous2023new, chow2023beware, fu2022vulrepair, mets2024automated, hartvigsen2022toxigen, botacin2023gpthreats, liu2023malicious}.

Despite the notable progress, security presents unique challenges when adopting language models. Universal challenges include hallucinations, context length limits, and the reasoning capabilities of language models. Additionally, security-specific challenges arise, such as dealing with logs and telemetry data, which often differ from natural language texts and are massive in volume.

Moreover, malicious threat actors (TAs) continuously exploit vulnerabilities to compromise systems and applications, aiming to harm or steal intellectual property. These actors vary in their motivations, with some driven by financial gain (e.g., ransomware groups) and others by political motives (e.g., state-sponsored actors). Their techniques also vary, ranging from complex social engineering to exploiting existing weaponized vulnerabilities.

This paper introduces \name{}, a novel approach for threat actor-informed prioritization of applications using SecEncoder, a security-specific language model \cite{stm_paper_2024}\footnote{This paper does not discuss the benefits of SecEncoder in detail. For more information, please refer to the original publication.}. Enterprise applications serve as crucial gateways for accessing vital resources within a company through mechanisms known as service accounts or principals. These principals theoretically have access to numerous resources necessary for the application’s operation, such as Key Vault, Emails, or Storage. Threat actors (TAs) often exploit this mechanism to compromise the service principal of an application, enabling lateral movement to other resources or information theft.

Whenever an action is performed on a resource, an event is logged in the resource’s databases, providing crucial breadcrumbs for monitoring and auditing suspicious activities. For service principals, sign-in logs are particularly important as they offer a gateway for accessing resources, providing critical context for identifying compromises. \name{} focuses on sign-in logs as the primary telemetry, while also utilizing other log types as necessary. Although the primary focus is on application investigation and risk-based prioritization using service principals, the concepts and methodologies are also applicable to other entities such as users or devices.

To achieve this objective, \name{} employs an encoder-only small security specific language model (\stm{}) alongside a decoder-only Large Language Model (GPT-4 Turbo) to analyze raw logs related to applications. This dual-model approach allows for a comprehensive analysis, determining if an application is compromised based on a given Threat Actor (TA) profile. Additionally, \name{} ranks the priority of investigations, helping security professionals focus on the most critical applications. By leveraging advanced language models, \name{} enhances the efficiency and effectiveness of security operations, ensuring that the most significant threats are addressed promptly.

Our contributions include:

\begin{itemize}

\item A novel architecture that combines encoder and decoder style language models, reducing the large volume of logs while maintaining diversity for effective reasoning.

\item Utilization of threat actor profiles to match application behavior, enabling effective identification of activities.

\item Experiments demonstrating the effectiveness of \name{} across various dimensions, including overall performance, separate results for malicious and benign applications, and detailed analysis of specific benign application cases.

\item \name{} effectively prioritizes applications, achieving an F-1 score of 0.90 in identifying malicious ones, and reduces analysts' workload by 87\%.



\end{itemize}

The rest of the paper is organized as follows: Section \ref{sec:related} reviews related work. Section \ref{sec:threat_model} outlines the threat model for applications, with a focus on Microsoft Azure as a cloud provider, though the concepts are broadly applicable to any cloud provider. Section \ref{sec:system} details \name{} and its overall design. Section \ref{sec:eval} presents the data, experiments, and results. Section \ref{sec:discussion} addresses limitations, discusses various experiences, and explores future work. Finally, Section \ref{sec:conclusion} concludes the paper.

\section{Related Work}
\label{sec:related}

Our work finds similarities to multiple lines of research. We discuss them below, highlighting the main differences between this and prior work, and the gaps we aim to fill.

Since \name{} essentially prioritizes applications for investigation, we first review existing cybersecurity risk assessment methods, reflecting on the frameworks and approaches used. We categorize these approaches into subjective, impact-focused, vulnerability-focused, and adversary-focused.

Subjective approaches rely on the assessor's expertise, often using frameworks like NIST SP 800-30, combined with interviews and questionnaires~\cite{lim2012risk, setiawan2017design, supriyadi2018information}. However, these can be limited by the assessor's knowledge. Impact-focused approaches prioritize critical assets and their potential impacts, using frameworks like ISO 27005~\cite{alwi2018information, harry2022effects}. These methods may not fully account for threat likelihood or actor behavior, but some studies aim to address these gaps (such as ~\cite{kure2019assets}). Vulnerability-focused approaches, such as \cite{george2019vulnerability}, \cite{bulut2022vuln} and \cite{aksu2017quantitative}, use metrics like CVSS and vulnerability scanners to assess risks, but tend to lack threat intelligence data.

Most relevant to our work is adversary-focused approaches, which focus on understanding threat actors, using frameworks like NIST SP800-30~\cite{katsumata2010cybersecurity, kure2019cyber, haji2019hybrid}. Some studies incorporate threat metrics and cyber kill chain concepts~\cite{hoffmann2020risk, figueira2020improving}, but do not fully integrate threat actor attributes. Prior work considered the MITRE ATT\&CK framework as a valuable resource for modeling cybersecurity threats~\cite{golushko2020application}, but its integration into risk assessment is often incomplete. We argue that, while there is significant research on cyber risk assessment, there is a need for more comprehensive approaches that consider behavioral metrics of threat actors and integrate frameworks like MITRE ATT\&CK more thoroughly.

Threat intelligence is also related to the problem studied in this paper. Research has highlighted issues with the coverage, timeliness, and accuracy of (open) threat intelligence (TI), including abuse feeds and blocklists~\cite{thomas2016abuse, metcalf2015blacklist}. Efforts to formalize and measure the quality of TI have introduced metrics for coverage, accuracy, timeliness, relevance, overlap, latency, and volume~\cite{li2019reading, schaberreiter2019quantitative, griffioen2020quality}, with studies on how to present TI quality to analysts~\cite{schlette2021measuring}.
The use of high-quality TI is also critical, with organizations facing challenges in interpreting TI, managing large volumes of information, and addressing false positives~\cite{ponemon}. Despite the operational issues similar to those of blacklists, TI offers high-level contextual information that could address these problems~\cite{bianco2019pyramid}. Nevertheless, a SANS survey in 2019 revealed that security operations analysts tend to value low-level indicators of compromise over high-level tactics, techniques, and procedures (TTPs), likely due to their focus on enriching alerts with technical details~\cite{brown2019evolution}. Our work takes advantage of both approaches, by enriching alerts with TTPs relevant to a treat actor.

Recent research has highlighted the potential of Large Language Models (LLMs) in addressing various cybersecurity challenges. LLMs have been effectively used in software security to detect vulnerabilities from descriptions and source code, and to create security patches and exploits, showing high accuracy in these tasks~\cite{charalambous2023new, chow2023beware, fu2022vulrepair}. They have also been applied to higher-level security tasks, such as analyzing security and privacy policies to classify documents and identify potential policy violations~\cite{mets2024automated, hartvigsen2022toxigen}. In network security, LLMs have proven capable of detecting and classifying different types of cyberattacks from network traffic data, including DDoS, port scanning, and botnet activities~\cite{alkhatib2022can, ali2023huntgpt, moskal2023llms}. For malware analysis, LLMs are being used to classify malware families and detect malicious domains and URLs through textual and behavioral analysis~\cite{botacin2023gpthreats, liu2023malicious}. Additionally, LLMs are aiding in social engineering defense by identifying phishing attempts via email content analysis~\cite{ranade2021generating, jamal2023improved}. Overall, LLMs are becoming an integral part of cybersecurity, offering the ability to process vast amounts of data and learn from it. We extend prior work by leveraging LLMs for threat actor informed application prioritization.

\begin{figure*}[h!]
    \centering 
    \includegraphics[width=1\textwidth]{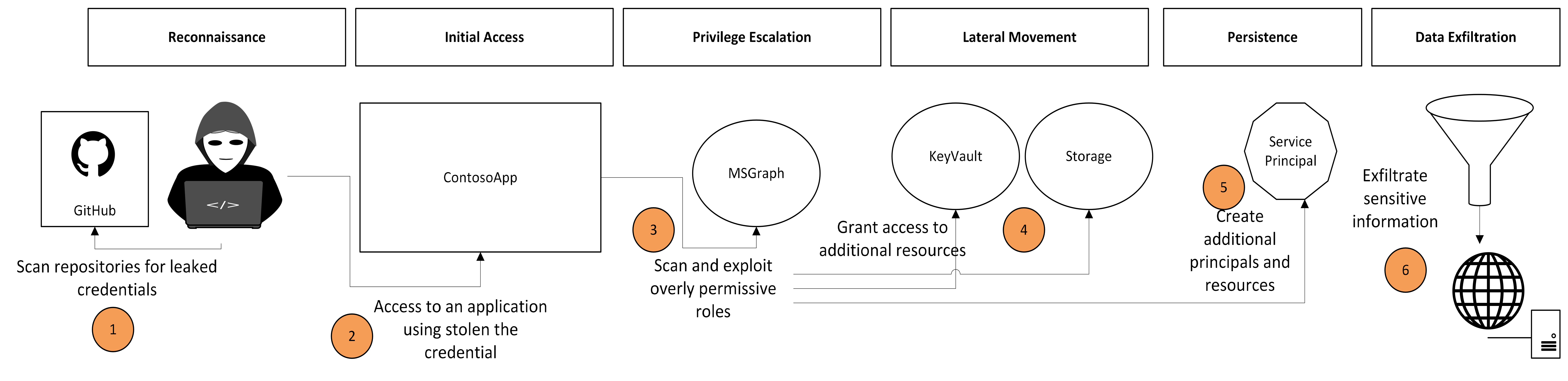}
    \caption{Steps to compromise an application.}
    \label{fig:threat_model}
\end{figure*}

\section{Threat Model for Compromising an Application}
\label{sec:threat_model}

In this section, we outline a potential threat model for compromising an application in Azure. While this model is specific to Azure, it is generally applicable to any cloud provider, such as AWS or Google Cloud, as they all support similar paradigms.

Figure \ref{fig:threat_model} illustrates a potential path through which an application can be compromised in Azure. Identity and access management (IAM) for a typical application in Azure is handled through Microsoft Entra \cite{microsoftEntra}, which serves as a gateway for authentication and authorization. The application itself may have an identity, known as a service principal, in addition to the users accessing the application. For the application to function properly, this service principal typically needs access to various resources such as Azure Key Vault, Azure Storage, or Microsoft Graph APIs. These accesses can be configured through the service principal with various levels of privileges.

A threat actor usually monitors public resources to seek vulnerabilities that can be exploited to gain access to an application (Reconnaissance). A forgotten credential on GitHub may provide such an opportunity for the actor to access the application. An expired credential may also leave traces in the form of certain error messages. Once the malicious actor gains access to the application via stolen or forgotten credentials, they may seek to escalate privileges by exploiting managed identities with overly permissive roles. If the application is configured with extensive privileges, this can potentially be abused to access additional resources and allow the threat actor to move laterally within the system. The threat actor can potentially create additional users and establish persistence for a longer presence. This presence would allow the malicious actor to exfiltrate data and sensitive information.

On the defensive side, when a user or service principal accesses the application, they must go through IAM, in this case, Entra, and leave a trace of activities that can be analyzed offline. Similarly, each resource, such as Key Vault, Storage, and Microsoft Graph, also maintains logs to track the activities of users and service principals. These logs are invaluable for detecting malicious activities and hunting for potential abuses. \name{} leverages these trails to identify threats and prioritize them for security analysts.

\section{System Design}
\label{sec:system}

\begin{figure*}[h!]
    \centering 
    \includegraphics[width=1\textwidth]{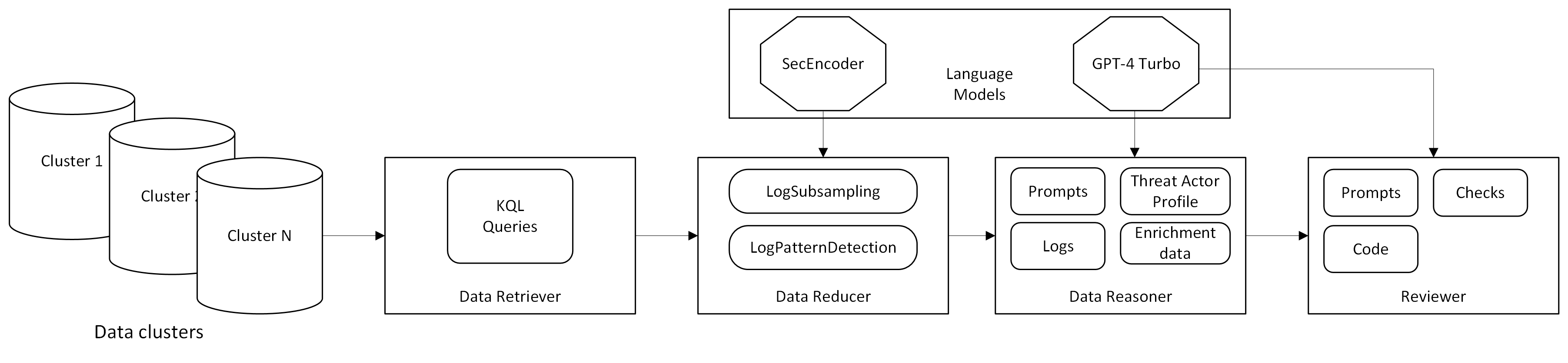}
    \caption{\name{} architecture, consisting of multiple components.}
    \label{fig:arch}
\end{figure*}

\name{} is designed with a modular architecture, consisting of five main components as shown in Figure \ref{fig:arch}: \emph{Language Models}, \emph{Data Retriever}, \emph{Data Reducer}, \emph{Data Reasoner}, and \emph{Reviewer}. The following sections explain each component in detail.

\subsection{Language Models}

\name{} relies on two state-of-the-art language models: the decoder-only GPT-4 Turbo model and the encoder-only \stm{} model \cite{stm_paper_2024}. GPT-4 Turbo is an enhanced version of the GPT-4 model, featuring several significant improvements. One of the most notable enhancements is its larger context window, which can handle up to 128,000 tokens, allowing it to process and retain more information in a single interaction. Additionally, GPT-4 Turbo is more cost-effective compared to its predecessor. These improvements make GPT-4 Turbo an excellent choice for working with logs and telemetries, as it can efficiently manage large volumes of data while maintaining high performance and cost efficiency.

\stm{} is a specialized small language model designed to address the unique challenges of cybersecurity. Unlike general-purpose language models, \stm{} is pretrained exclusively on a large corpus of security logs, capturing various events and activities related to security incidents and operations. This domain-specific pretraining enables \stm{} to outperform traditional language models, such as BERT and OpenAI's embedding model (text-embedding-ada-002), in tasks related to security log analysis \cite{stm_paper_2024}.

\stm{} employs an encoder-only architecture based on the DeBERTa-v2 model, which enhances BERT by incorporating a disentangled attention mechanism and an improved mask decoder. The model is pretrained using a customized masked language modeling (MLM) loss that focuses on learning the content of security logs. The training dataset comprises a diverse set of logs from both public and private sources, totaling approximately 1 terabyte of data. This extensive pretraining allows \stm{} to generate high-quality representations of security logs, making it a powerful tool for various security use cases, including log analysis, anomaly detection, and log search.

\subsection{Data Retriever}
This component is responsible for retrieving relevant telemetry and details about the application. The data resides in Kusto clusters, and Kusto Query Language (KQL) queries are used to fetch this data. The data retrieval process works in multiple steps.

In the first step, the component pulls the sign-in logs, which serve as the primary entry point for any subsequent activities. Next, it checks for access to other resources such as Microsoft Graph (MSGraph) or Azure Key Vault. If access to these resources is detected, relevant telemetry from those resources is also fetched, even if they reside in different clusters.

The observed IP addresses from these resources are then extracted, combined, and enriched with various details, including city, ISP, and proxy information. Additionally, other application-related data is retrieved, particularly the application’s permission details, such as which resources it has access to, and credential details, including creation, rotation, and expiration dates. Relevant alerts and incidents from other detectors, if they exist, are also gathered.

All of this data is used by downstream components to analyze the application for potential malicious activities. This comprehensive approach ensures that any suspicious behavior is detected and addressed promptly, enhancing the overall security posture of the application.

\subsection{Data Reducer}
This component addresses the critical need to reduce data volume. Some applications generate an excessive amount of logs that cannot be accommodated within the LLM prompt, rendering them unusable directly. To overcome this challenge, \name{} employs two novel approaches from \stm{}: LogSubsampling and LogPatternDetection \cite{stm_paper_2024}.

Logs that exceed certain size thresholds are processed by the Data Reducer components. These innovative methods enable \name{} to retain relevant data while eliminating unnecessary noise, thereby focusing on the most critical information.

LogSubsampling selectively reduces the volume of log data by sampling key entries that are representative of the overall dataset. This ensures that significant events and patterns are preserved without overwhelming the system with redundant information.

LogPatternDetection, on the other hand, identifies and extracts anomalies within the logs. By recognizing these patterns, the system can condense the data into a more manageable form, highlighting essential activities and anomalies that require attention.

Together, these approaches allow \name{} to efficiently manage large volumes of log data, ensuring that the most pertinent information is available for analysis. This not only enhances the performance of the system but also improves the accuracy and relevance of the insights derived from the data. Following sections explains the methodologies in details.

\subsubsection{LogSubsampling} 
Logs are a rich source of security data, providing valuable insights into system activities and potential threats. However, security analysts face significant challenges when investigating large quantities of raw logs and transforming them into actionable insights and recommendations. Generative models like GPT can assist in this transformation by turning raw data into meaningful insights and recommendations. However, due to size limitations, language models cannot effectively process vast amounts of raw logs in their entirety.

While generic methods exist for breaking large datasets into smaller chunks and combining results, these methods often fail to account for the varying importance of different data chunks. This is where LogSubsampling, powered by \stm{} embeddings, comes into play. LogSubsampling effectively selects chunks of data that retain the critical information needed for an investigation, while pruning away redundant and uninformative data. This ensures that the most relevant and diverse set of logs is chosen for analysis.

For subsampling, \stm{} employs a greedy algorithm based on \stm{} embeddings. This algorithm selects logs that maximize the variance of the embeddings, ensuring the chosen logs are both diverse and informative. Initially, the algorithm selects the log closest to the center of all logs. It then iteratively selects the log with the maximum minimum distance to the embeddings of the already selected logs. This \emph{Max-Min} strategy ensures the selected logs are maximally diverse and avoids choosing logs that are too similar to those already selected.

\subsubsection{LogPatternDetection}
LogPatternDetection is a tool designed to identify anomalies in log data. It employs an unsupervised implementation of the IsolationForest algorithm, utilizing \stm{} embeddings as a featurizer. 

The IsolationForest algorithm operates by creating a random subsample of the dataset, which is then used to construct an isolation tree. Each tree is built using \stm{} embedding as a feature, with split values selected randomly. The resulting binary tree is analyzed to identify isolated points, which are expected to have fewer splits, indicating potential anomalies. This method is effective for anomaly detection in logs because it does not require labeled data and can handle high-dimensional datasets efficiently.

\subsection{Data Reasoner}

This component utilizes condensed data from the Data Reducer, along with other application-specific details, to analyze activities and determine a triage priority level. It relies on a specified Threat Actor (TA) profile, which can range from generic MITRE ATT\&CK techniques to detailed subtechniques. The Data Reasoner uses this profile to match against the observed activities. Appendix \ref{sec:app:profile} provides an example of a TA profile defined in YAML format.

The primary reasoning engine for \name{} is a Large Language Model (LLM), specifically GPT-4 Turbo, which features an extensive context window of 128,000 tokens. This capability is crucial for processing raw logs and other extensive data inputs. Next, we show an example of a system prompt used in this context.

\begin{tcolorbox}[colframe=black, width=0.5\textwidth, arc=0mm, outer arc=0mm, title=Example system prompt, label={box:system_prompt1}]
\small
As a cybersecurity expert, you specialize in investingating application logs and find signs of compromise. You understand MITRE ATT\&CK Matrix, Cyber Kill Chain and TTPs (Techniques, Tactics and Practices). Your organization is under attack by a Threat Actor. You will be given the profile of the threat actor in terms of Cyber Kill Chain stages and TTPs used by the threat actor. Your task is to analyze application logs and find signs of compromise and malicious activities that match the threat actor profile.
\end{tcolorbox}

While the system prompt is designed to remind the LLM of its role in \name{}, the user prompt is meticulously crafted to include all necessary details, enabling the LLM to make informed decisions. The prompt begins with a section that outlines the information and instructions the LLM should expect, explicitly highlighting critical details such as the Threat Actor Profile. This section ensures that the LLM is aware of the context and the specific threat landscape it needs to consider. Following this, the prompt provides guidance on how to interpret the data, explaining the different types of telemetry and data sources involved. This helps the LLM understand the nature of the data it will be processing and the significance of various data points. The next part of the prompt details the expected output, which is divided into three main sections:

\begin{itemize}
\item High-Level Behavior of the Application: This section summarizes the overall behavior and patterns observed in the application.
\item Suspicious Activities and Entities Involved: Here, the LLM identifies any activities or entities that appear suspicious based on the provided data and threat profile.
\item Triage Priority Level: Based on the guidance provided, the LLM assigns a triage priority level to the application, indicating the urgency and severity of the potential threat.
\end{itemize}

Finally, the actual data and threat profile are included in the prompt. This approach ensures that the LLM has all the information it needs to perform its analysis effectively. The user prompt is structured to be flexible, allowing for the omission of any information that is missing or unknown. Additionally, it supports the inclusion of certain details as configuration files, such as guidance, data types, and the TA profile. This modular design enhances the prompt’s adaptability and ensures that it can be tailored to different scenarios and requirements. Next, we provide an example user prompt.


\begin{tcolorbox}[colframe=black, width=0.5\textwidth, title=Example user prompt, label={box:user_prompt1}]
\small
You are presented with various types of logs from an application that might have been compromised by the Threat Actor. The logs span a time period when the Threat Actor might have performed malicious activities. However, the logs may also contain benign activities that are not related to the compromise. \\
            
You will be given the '\# Threat Actor Profile'. The profile consists of a short description of the threat actor, followed by MITRE ATT\&CK Matrix consisting of Cyber Kill Chain stages. Under each Kill Chain stage you will find TTPs (Techniques, Tactics and Practices) used by the threat actor during the attack.\\
            
Your task is to analyze application logs using the supplemental enrichment data and find suspicious activity matching threat actor profile. You are going to list the Kill Chain Stages. Under each stage you'll list the matching TTP whose evidence you have found in the logs.\\
            
Consider the following guidance before judging an activity as suspicious: \\
\hspace*{2em} \{GUIDANCES\}\\

You will be given the '\# Application Logs' to help you analyze the activity. Application logs consist of the following logs:\\
\hspace*{2em} \{LOG\_TYPES\}\\

You will also be given the '\# Enrichment Data' to help you analyze the logs. Enrichment data consists of the following data types:\\
\hspace*{2em}  \{ENRICHMENT\_DATA\_TYPES\}\\

You should focus on all the data, while keeping in mind that there might be benign activities. If you require additional information, you should clearly indicate it in your response. Your response should include: \\
\hspace*{2em}             - High level behavior of the application.\\
\hspace*{2em}             - Suspicious activities and entities involved.\\
\hspace*{2em}             - Triage priority level of the application. Use the following guidances that include Kill Chain stages observed and other guidance. In your output you MUST list all of the guidance and whether the statement is true or false. Add "[True]" or "[False]" at the end of each guidance.\\
\hspace*{4em}			\{GUIDANCES\}\\

\{INPUT\_DATA\}
\end{tcolorbox}

\subsection{Reviewer}
Large Language Models (LLMs) are generally effective in analyzing given data and arriving at conclusions. However, it has been observed that as the context provided to an LLM increases, its effectiveness can diminish, leading to overlooked details. This is a known limitation of state-of-the-art LLMs \cite{liu2023lostmiddlelanguagemodels}.

To mitigate this limitation, \name{} employs a secondary step to ensure that no critical details are missed. This is achieved through a reviewer component, which utilizes natural language-defined \emph{checks} and \emph{codes} to verify the output of the Data Reasoner.

Checks include guidelines such as double-checking whether certain IP addresses are correctly classified as benign based on specific criteria. These checks ensure that the initial analysis by the LLM is accurate and comprehensive.

Codes involve the use of regular expressions and other programming constructs to validate and verify the report. For instance, regular expressions can be used to ensure that the report highlights high-priority sensitive resources. These codes act as an additional layer of scrutiny to catch any details that might have been missed by the LLM.

The \emph{Reviewer} component ensures that the final output is robust and reliable. By incorporating both checks and codes, it validates the findings of the \emph{Data Reasoner}, ensuring that the report is thorough and can be confidently used in the downstream triage process.

This two-step approach enhances the reliability of \name{}'s analysis by addressing the inherent limitations of LLMs when dealing with large contexts, ensuring that critical details are not overlooked. Appendix \ref{sec:app:output} provides an example output from \name{}.

\section{Evaluation}
\label{sec:eval}

This section provides a comprehensive discussion on the data, detailing the methodology used for evaluation and the experiments conducted.

\subsection{Dataset}
We have curated 93 applications, 32 of them are malicious and 61 are benign apps. We further divide the benign apps into two categories: benign-nonsuspicious 16 apps and benign-suspicious 45 apps. Definitions of the different categories are described in Table \ref{tab:app_categories}.

\begin{table}[t]
	\centering
	\begin{adjustbox}{width=0.44\textwidth,keepaspectratio}
		\begin{tabular}{c l L{4cm}}
			\textbf{Code} & \textbf{Category} & \textbf{Definition}\\
			\cmidrule(lr){1-1} \cmidrule(lr){2-2} \cmidrule(lr){3-3}
			0 & Benign-nonsuspicious & Applications with no suspicious behavior\\[5mm]
			1 & Benign-suspicious & Applications with suspicious behaviors that require analyst examination such as unsuccessful attempts, tricky developer behavior, etc.\\[5mm]
			2 & Compromised & Applications for which the TA has performed malicious activities with the existing privileges of the application
		\end{tabular}
	\end{adjustbox}
	\caption{Application Categories and Definitions.}
	\label{tab:app_categories}
\end{table}

\begin{figure}[h!]
    \centering 
    \includegraphics[width=0.5\textwidth]{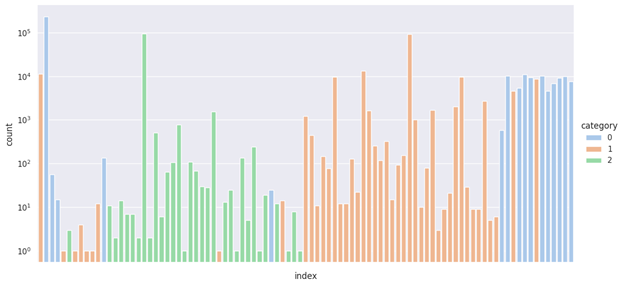}
    \caption{Sign-in counts. Each bar represents one application. Y axis in logarithmic scale.}
    \label{fig:app_signin_count}
\end{figure}

Each application is analyzed over a period of approximately three months. During this time, the distribution of sign-in counts for these applications is shown in Figure \ref{fig:app_signin_count}, displayed on a logarithmic scale to account for some applications with over 200,000 sign-in records.

\subsection{Methodology}
\name{} assigns a priority score for each application based on a comprehensive list of criteria. These criteria are detailed in Table \ref{tab:criteria_scores_v0.7} and Table \ref{tab:criteria_scores_v0.4}, which reflect two distinct sets of criteria, differing in both breadth and scope. 
The first set (\baseline{}) constitutes our baseline, which encompasses a wide range of MITRE ATT\&CK tactics, providing an extensive evaluation framework.
In contrast, the second set (\focused{}) is more concise and specifically tailored to the activities of the Threat Actor (TA), focusing on targeted aspects of their behavior.

\begin{table}[t]
	\centering
	\begin{adjustbox}{width=0.49\textwidth,keepaspectratio}
		\begin{tabular}{l L{6.5cm} L{1.25cm}}
			\textbf{Source} & \textbf{Criteria} & \textbf{Score $\Delta$}\\
			\cmidrule(lr){1-1} \cmidrule(lr){2-2} \cmidrule(lr){3-3}
			TA Profile & Initial Access observed & +1\\
			TA Profile & Execution observed & +1\\
			TA Profile & Persistence observed. & +1\\
			TA Profile & Reconnaissance observed & +1\\
			TA Profile & Privilege Escalation observed & +1\\
			TA Profile & Defense Evasion observed & +1\\
			TA Profile & Credential Access observed & +1\\
			TA Profile & Lateral Movement observed & +1\\
			TA Profile & Data Collection observed & +1\\
			Other & Application has extensive permissions, can access sensitive resources & +1\\
			Other & More than one stages of kill chain observed & +1\\
			Other & All access attempts were unsuccessful and resulted in errors & -1\\
			Other & Application lacks access sensitive resources & -1\\
			Other & All the suspicious IP addresses are benign & -3\\
			Other & All the suspicious IP addresses were linked to resources not considered high & -1\\
		\end{tabular}
	\end{adjustbox}
	\caption{\baseline{} priority score criteria for an application and corresponding score contributions.}
	\label{tab:criteria_scores_v0.7}
\end{table}

\begin{table}[t]
	\centering
	\begin{adjustbox}{width=0.49\textwidth,keepaspectratio}
		\begin{tabular}{l L{6.5cm} L{1.25cm}}
			\textbf{Source} & \textbf{Criteria} & \textbf{Score $\Delta$}\\
			\cmidrule(lr){1-1} \cmidrule(lr){2-2} \cmidrule(lr){3-3}
			TA Profile & Unusual Access Attempt observed & +1\\
			TA Profile & OAuth Abuse observed & +2\\
			TA Profile & Data Collection Activities observed & +1 \\
			TA Profile & Use of Proxy Infrastructure observed & +1\\
			Other & Application has extensive permissions, can access sensitive resources & +1\\
			Other & More than one method of threat actor observed & +1\\
			Other & A suspicious pattern of accessing sensitive resources observed & +1\\
			Other & All access attempts were unsuccessful and resulted in errors & -1\\
			Other & Application lacks access to sensitive resources & -1\\
			Other & All the suspicious IP addresses are benign & -2\\
			Other & All the suspicious IP addresses were linked to resources not considered high & -1\\
		\end{tabular}
	\end{adjustbox}
	\caption{\focused{} priority score criteria for an application and corresponding score contributions.}
	\label{tab:criteria_scores_v0.4}
\end{table}

These criteria have been meticulously defined based on the extensive experience and domain knowledge of security analysts. They represent the key items that analysts would manually evaluate in each application. Each criterion contributes a specific score, which is summed to determine the overall priority score for the application. The maximum score an application can receive is 11 (or 8), while the minimum score is 0. The scores are extracted from \name{}'s output, particularly from Triage Priority Level section.

Given that some applications contain more than 200,000 records, our analysis divides them into multiple segments based on the limitations of data retrieval from Kusto clusters, which is approximately 40,000 records at a time. For instance, if an application has 80,000 records, it will be divided into two analysis periods. The representative score for the application is the highest score obtained during these periods.

To account for variability in scoring by the Language Model (LLM), each application is evaluated five times. The representative score for the application is determined through majority voting. Based on a chosen threshold of priority score, the application is classified as malicious or benign. We use the following metrics to evaluate \name{}:

\begin{itemize}[noitemsep]
\item \textbf{Precision}: This measures the accuracy of the positive predictions. It is the ratio of true positive results to the total predicted positives. High precision indicates that the model has a low false positive rate.

\item \textbf{Recall}: This measures the ability of the model to find all the relevant cases within a dataset. It is the ratio of true positive results to the total actual positives. High recall indicates that the model has a low false negative rate.

\item \textbf{F1-score}: This is the harmonic mean of precision and recall. It provides a single metric that balances both precision and recall, especially useful when you need to find an optimal balance between the two.
\end{itemize}

Since there are two labels (benign and malicious applications), we calculate the metrics using a weighted average, which ensures that each label’s contribution is proportional to its occurrence in the dataset. More specifically, each metric for a label is multiplied by the number of instances of that label, and then the sum of these products is divided by the total number of instances across all labels.

To compare different methods, such as \baseline{} and \focused{}, we use three minimum priority scores--3, 4 and 5--as thresholds and take the average as the representative metric for each methodology. For example, if the minimum priority score is defined as 3, then any score of 3 or above classifies the application as malicious.

\subsection{Experiments}

We evaluate \name{} across multiple dimensions. In Section \ref{subsub:overall}, we discuss the overall performance of \name{} and provide a comparative analysis between versions \baseline{} and \focused{}. Section \ref{subsub:malicious} delves into the performance specifics for malicious applications, highlighting key metrics and observations. Section \ref{subsub:benign} focuses on benign applications, with an emphasis on precision and recall metrics. Further, Section \ref{subsub:benign_details} explores the performance metrics in greater detail for various benign categories, as outlined in Table \ref{tab:app_categories}. 


\subsubsection{Overall Performance}
\label{subsub:overall}

\begin{figure*}[h!]
    \centering
    \begin{subfigure}[b]{0.5\textwidth}
        \centering
        \includegraphics[width=\textwidth]{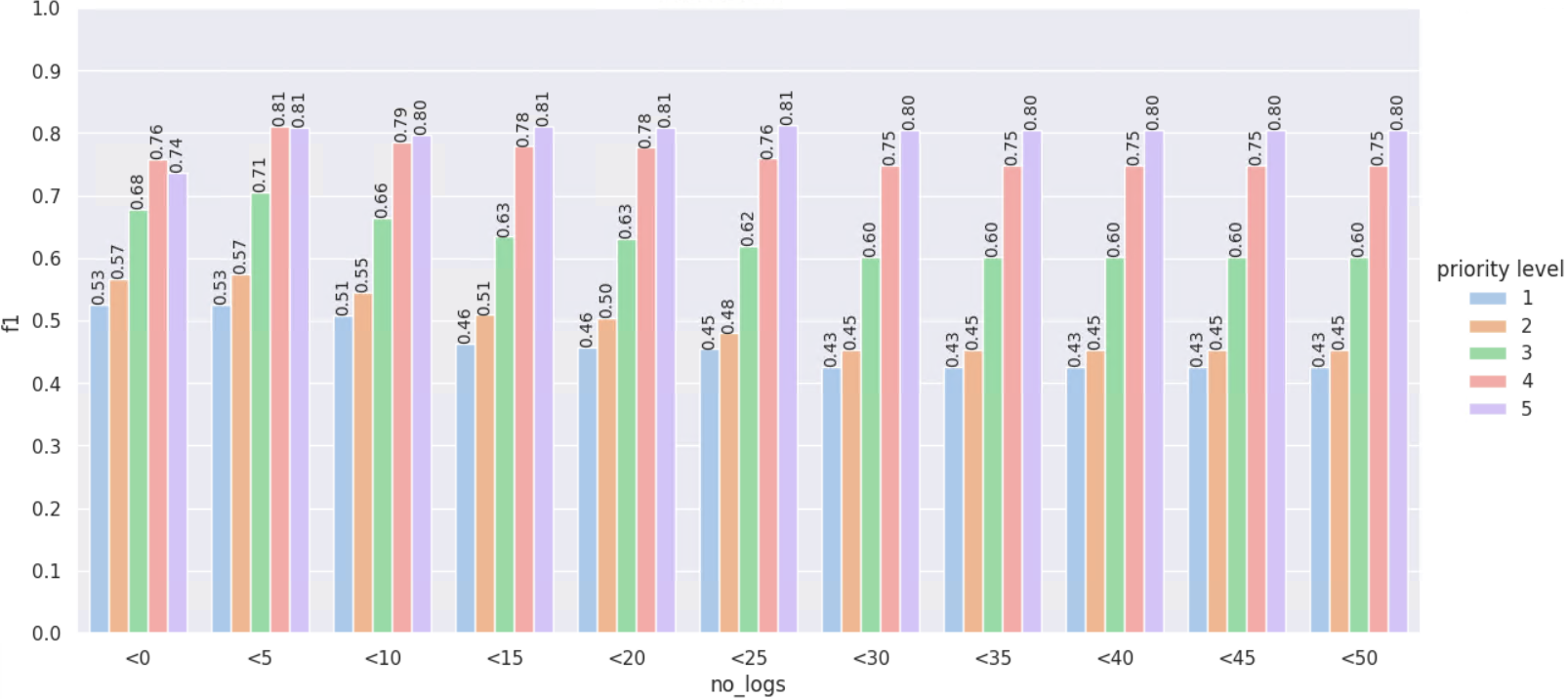}
        \caption{Overall performance of \baseline{} using F-1 score.}
        \label{fig:overall_v0.7}
    \end{subfigure}%
    \begin{subfigure}[b]{0.5\textwidth}
        \centering
        \includegraphics[width=\textwidth]{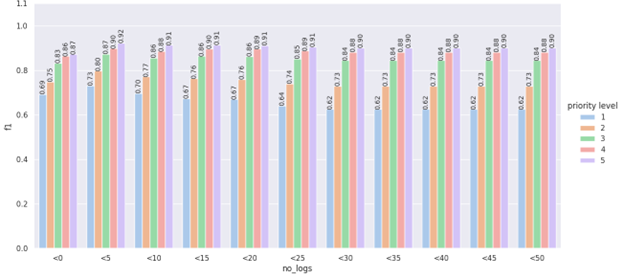}
        \caption{Overall performance of \focused{} using F-1 score.}
        \label{fig:overall_v0.4}
    \end{subfigure}
    \caption{Comparison of \baseline{} and \focused{} using F-1 score.}
    \label{fig:overall_comparison}
\end{figure*}

Figure \ref{fig:overall_comparison} illustrates the overall performance of \name{} using the F1-score metric. Each bar group on the x-axis represents the count of sign-in logs greater than the specified number. For instance, the group labeled “<5” indicates performance for applications with at least 5 or more sign-in log counts. Each bar, distinguished by different colors, represents various minimum thresholds of priority scores used for predicting whether an application is malicious or benign.

From Figure \ref{fig:overall_comparison}, it is evident that there is a slight increase in performance as the required number of log counts rises from 0 to at least 5. This trend is expected, as a higher number of logs provides more context for the application, enabling \name{} to make more informed decisions. When considering a minimum of 5 logs for an application and averaging the performance for priority scores of 3, 4, and 5 or above as malicious, \focused{} achieves an F1-score of 0.90, whereas \baseline{} achieves an F1-score of 0.78.

We attribute this difference to the fact that \focused{} employs a more targeted and straightforward priority calculation list compared to \baseline{}. However, this advantage comes with a trade-off: \focused{} requires more prior knowledge, which may not be suitable for more complex investigations. Despite this, the higher F1-score of \focused{} indicates its effectiveness in scenarios where sufficient prior knowledge is available.

\subsubsection{Malicious Applications Performance Details}
\label{subsub:malicious}

\begin{figure*}[h!]
	\centering
	\begin{subfigure}[b]{0.5\textwidth}
		\centering
		\includegraphics[width=\textwidth]{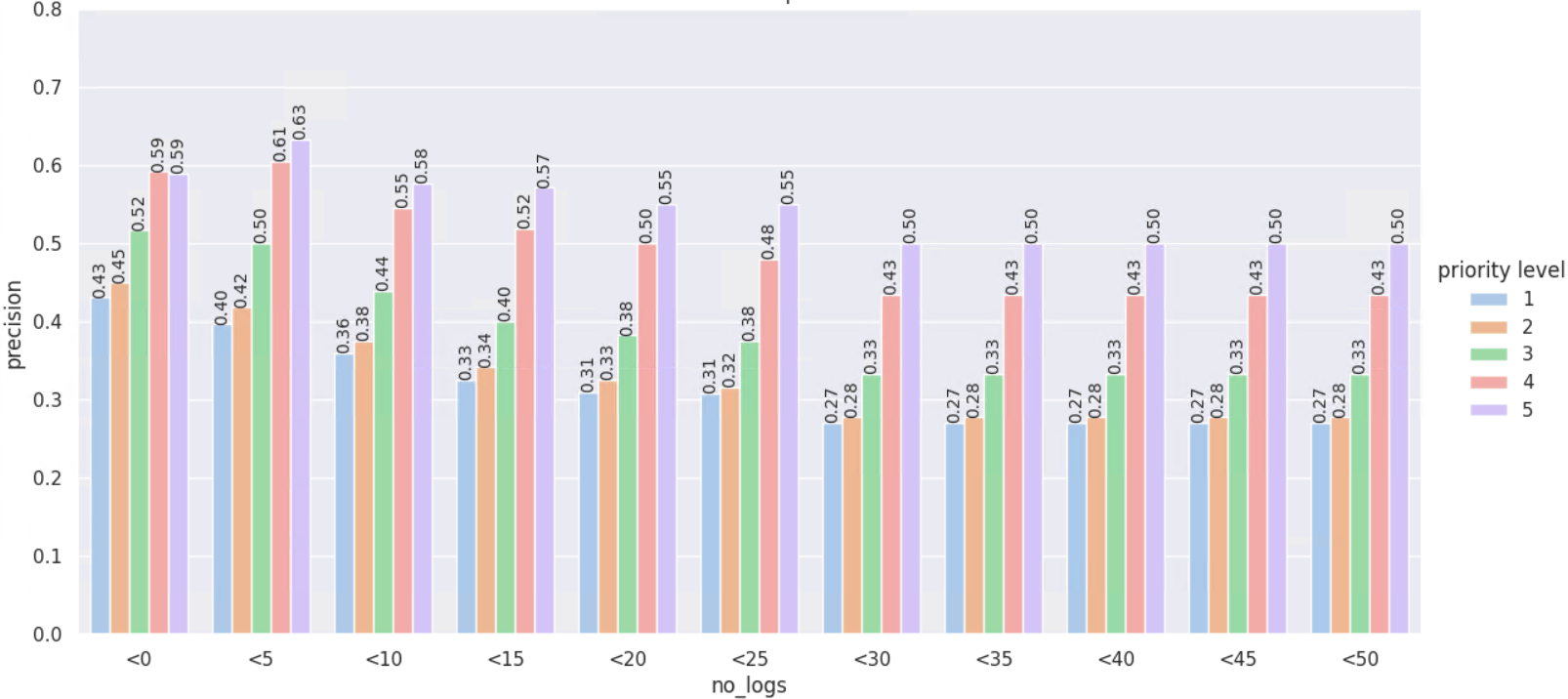}
		\caption{Precision for \baseline{} for malicious apps.}
		\label{fig:malicious_precision_v0.7}
	\end{subfigure}%
	\begin{subfigure}[b]{0.5\textwidth}
		\centering
		\includegraphics[width=\textwidth]{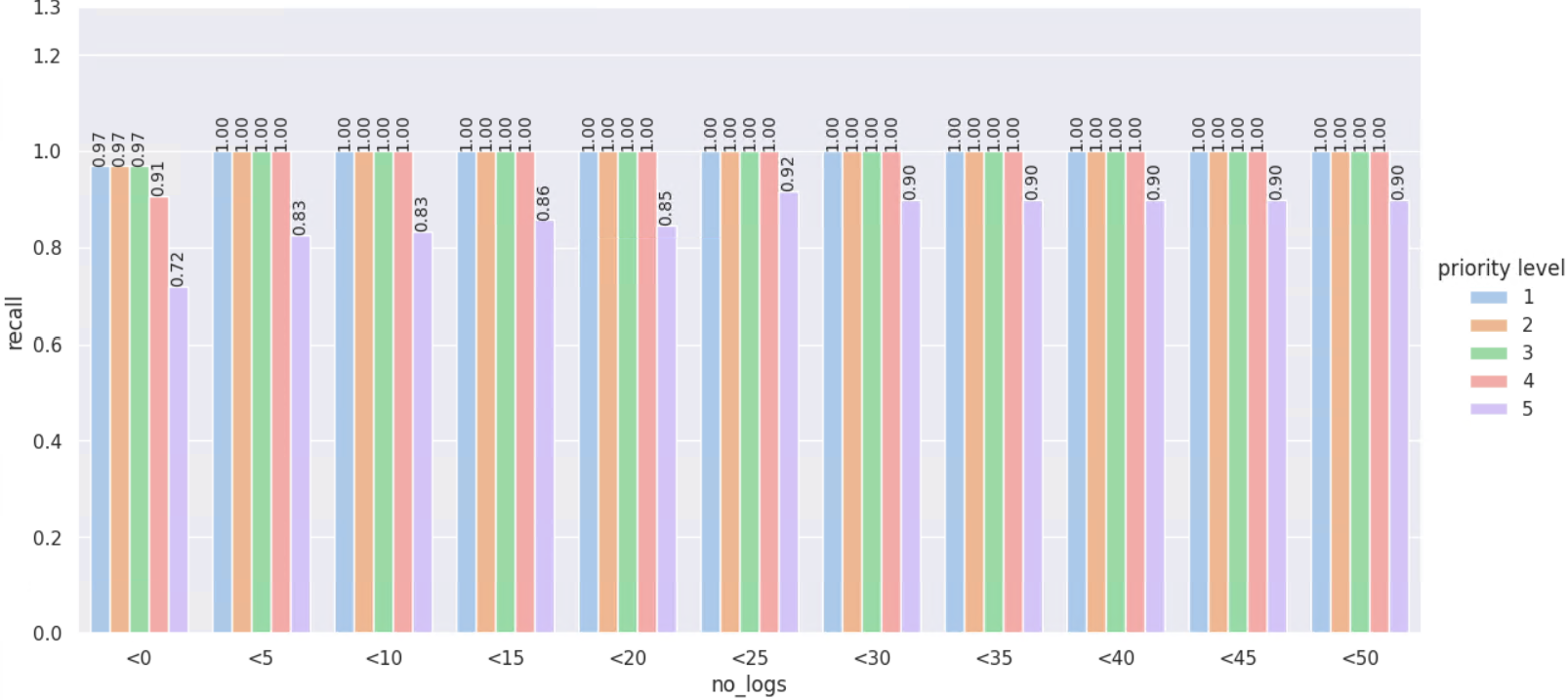}
		\caption{Recall for \baseline{} for malicious apps.}
		\label{fig:malicious_recall_v0.7}
	\end{subfigure}
	\caption{Comparison of precision and recall for \baseline{} for malicious apps.}
	\label{fig:malicious_comparison_v0.7}
\end{figure*}

\begin{figure*}[h!]
    \centering
    \begin{subfigure}[b]{0.5\textwidth}
        \centering
        \includegraphics[width=\textwidth]{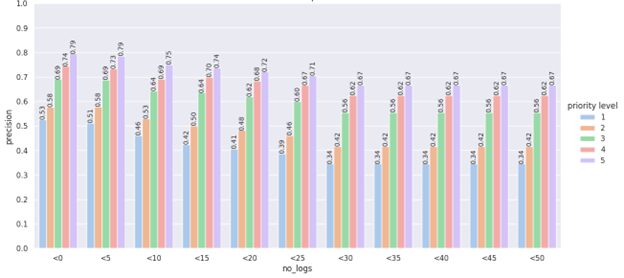}
        \caption{Precision for \focused{} for malicious apps.}
        \label{fig:malicious_precision_v0.4}
    \end{subfigure}%
    \begin{subfigure}[b]{0.5\textwidth}
        \centering
        \includegraphics[width=\textwidth]{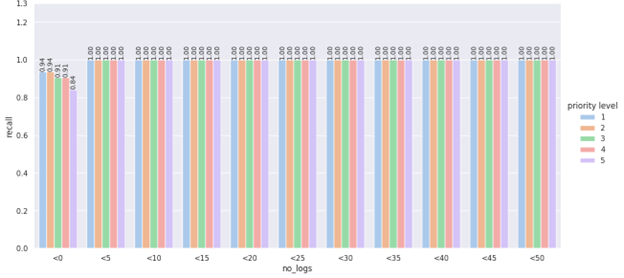}
        \caption{Recall for \focused{} for malicious apps.}
        \label{fig:malicious_recall_v0.4}
    \end{subfigure}
    \caption{Comparison of precision and recall for \focused{} for malicious applications.}
    \label{fig:malicious_comparison_v0.4}
\end{figure*}

Figures \ref{fig:malicious_comparison_v0.7} and \ref{fig:malicious_comparison_v0.4} illustrate the precision and recall metrics for identifying malicious applications for \baseline{} and \focused{}. By averaging the priority scores of 3, 4, and 5, with a minimum of 5 logs, we achieve a precision of 0.74 and a recall of 1.00 for \focused{}, and a precision of 0.58 and a recall of 0.94 for \baseline{}. Lowering the priority level threshold consistently introduces more false positives in both versions.

In both versions, \name{} consistently achieves high recall, demonstrating its strength in not missing actual malicious applications. As discussed in the previous section, the shorter and more targeted priority criteria of \focused{} outperformed the longer and broader priority criteria of \baseline{}. This indicates that a more focused approach is more effective in accurately identifying malicious applications while maintaining a high recall rate.

\subsubsection{Benign Application Performance Details}
\label{subsub:benign}

\begin{figure*}[h!]
	\centering
	\begin{subfigure}[b]{0.5\textwidth}
		\centering
		\includegraphics[width=\textwidth]{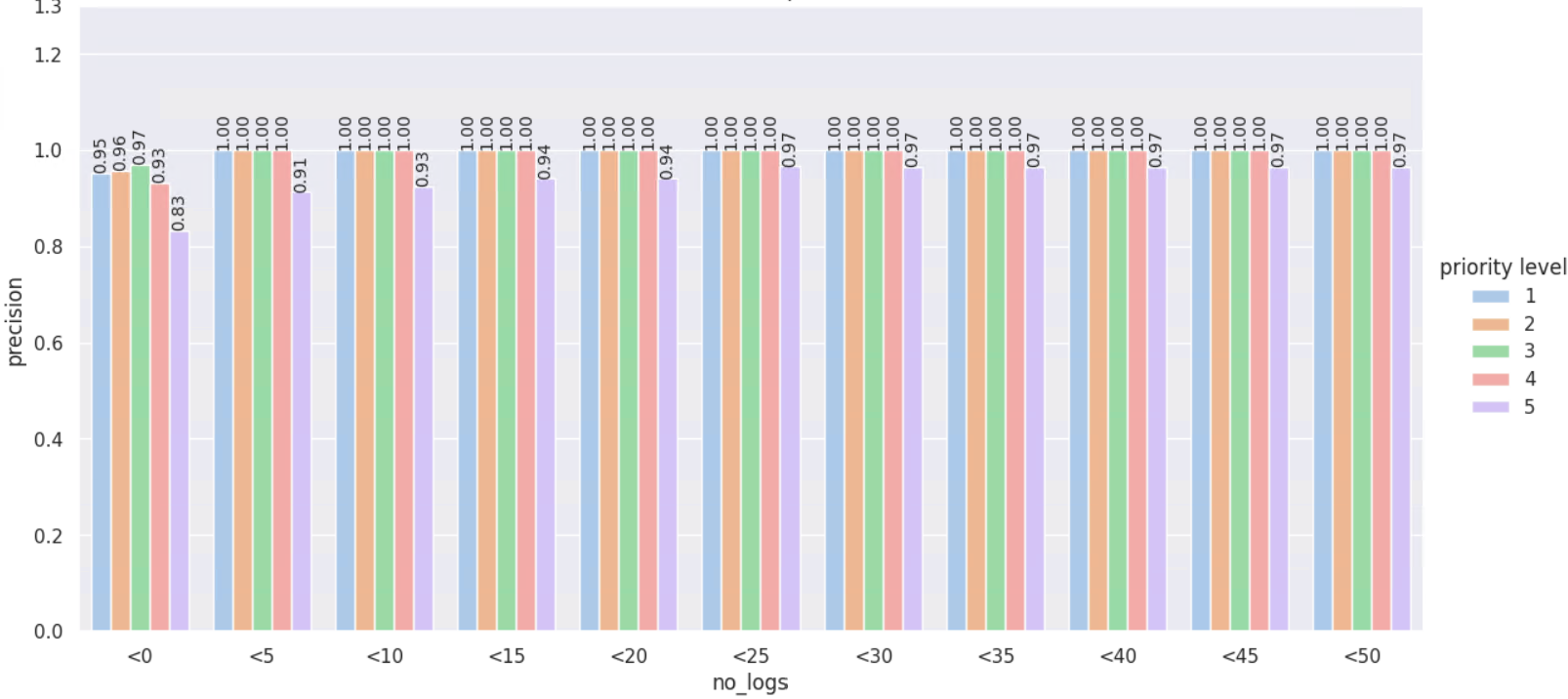}
		\caption{Precision for \baseline{} for benign apps.}
		\label{fig:benign_precision_v0.7}
	\end{subfigure}%
	\begin{subfigure}[b]{0.5\textwidth}
		\centering
		\includegraphics[width=\textwidth]{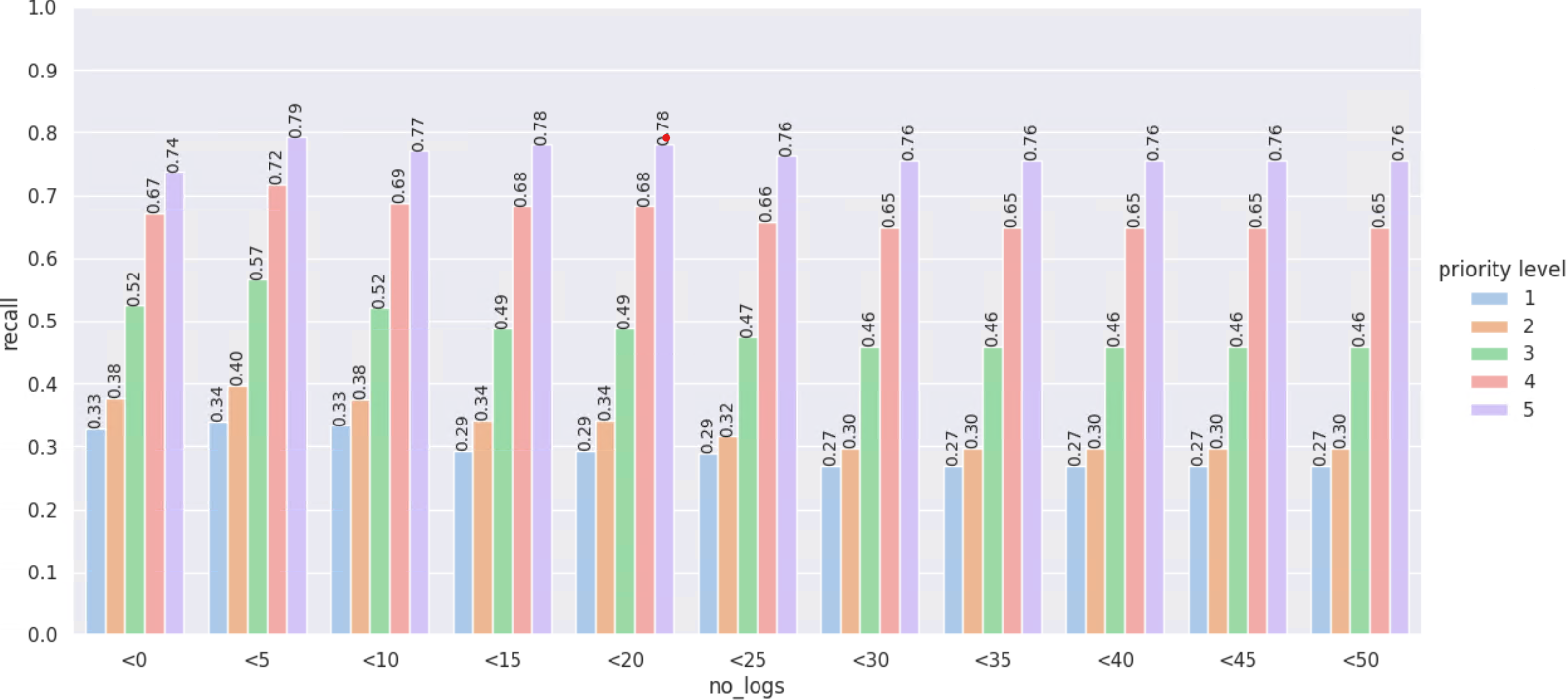}
		\caption{Recall for \baseline{} for benign apps.}
		\label{fig:benign_recall_v0.7}
	\end{subfigure}
	\caption{Comparison of precision and recall for \baseline{} for benign apps.}
	\label{fig:benign_comparison_v0.7}
\end{figure*}

\begin{figure*}[h!]
    \centering
    \begin{subfigure}[b]{0.5\textwidth}
        \centering
        \includegraphics[width=\textwidth]{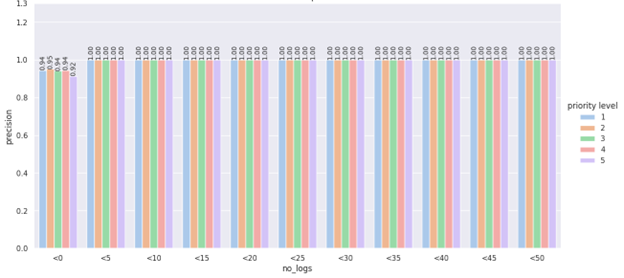}
        \caption{Precision for \focused{} for benign apps.}
        \label{fig:benign_precision_v0.4}
    \end{subfigure}%
    \begin{subfigure}[b]{0.5\textwidth}
        \centering
        \includegraphics[width=\textwidth]{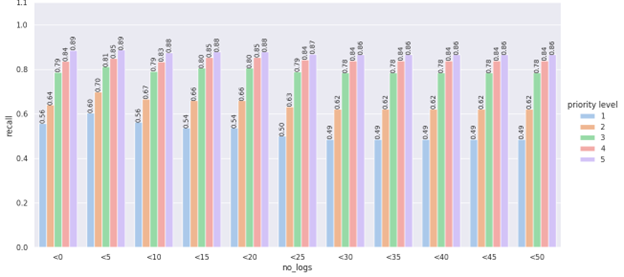}
        \caption{Recall for \focused{} for benign apps.}
        \label{fig:benign_recall_v0.4}
    \end{subfigure}
    \caption{Comparison of precision and recall for \focused{} for benign apps.}
    \label{fig:benign_comparison_v0.4}
\end{figure*}

Figures \ref{fig:benign_comparison_v0.7} and \ref{fig:benign_comparison_v0.4} illustrate the precision and recall metrics for benign applications using \baseline{} and \focused{}. By averaging the priority scores of 3, 4, and 5, with a minimum of 5 logs, \focused{} achieves a precision of 1.0 and a recall of 0.85. In comparison, version \baseline{} achieves a precision of 0.97 and a recall of 0.69.

The near-perfect precision of \name{} allows security analysts to confidently ignore applications classified as benign, significantly reducing their workload. This efficiency is further demonstrated in an internal real-world study, where 412 suspicious applications were analyzed by \name{}. Randomly selected applications classified as benign by \name{} were picked and verified for accuracy by security analysts. All of the randomly selected applications were indeed benign, resulting in an approximately 87\% reduction in the number of applications that analysts needed to investigate. This substantial decrease in the analysts’ queue translates to a more streamlined and efficient security process, enabling analysts to focus on more critical tasks.

\subsubsection{Different Categories of Benign Applications}
\label{subsub:benign_details}

In this section, we investigate whether different types of benign applications pose any challenges for \name{}. Our goal is to determine if the diversity in benign application types affects the performance and accuracy of \name{} in classifying them correctly. By analyzing various categories of benign applications, we aim to identify any potential difficulties or inconsistencies that may arise. This analysis will help us understand the robustness of \name{} and its ability to maintain high recall across a wide range of benign application types. Since this analysis is purely focusing on benign apps, we use recall as a metric to compare performances.

\begin{figure*}[h!]
    \centering
    \begin{subfigure}[b]{0.5\textwidth}
        \centering
		\includegraphics[width=\textwidth]{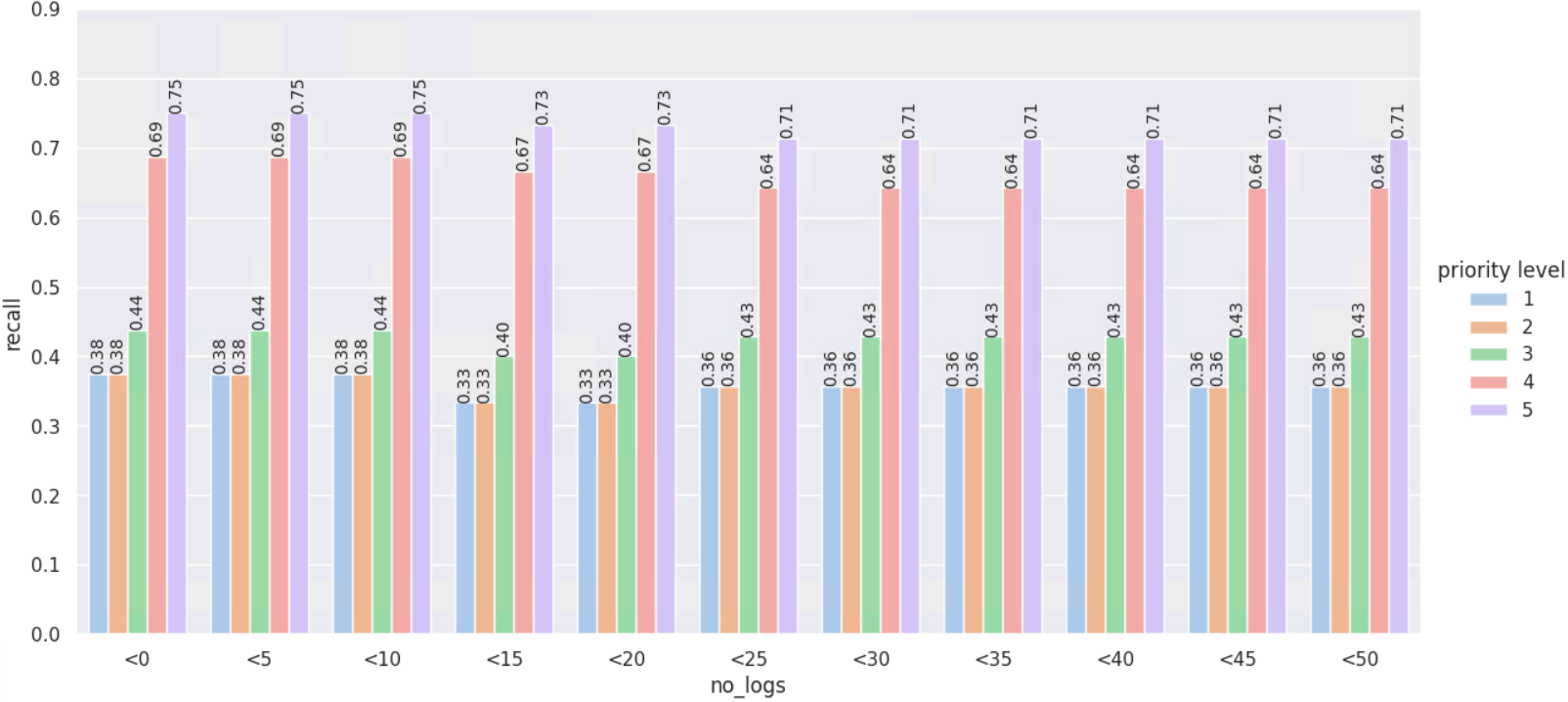}
        \caption{Recall for \baseline{} for benign-nonsuspicious apps.}
        \label{fig:benign_ns_recall_v0.7}
    \end{subfigure}%
    \begin{subfigure}[b]{0.5\textwidth}
        \centering        
        \includegraphics[width=\textwidth]{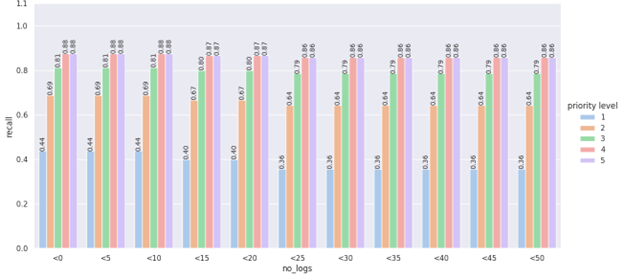}
        \caption{Recall for \focused{} for benign-nonsuspicious apps.}
        \label{fig:benign_ns_recall_v0.4}
    \end{subfigure}
    \caption{Comparison of recall for \baseline{} and \focused{} for benign-nonsuspicious apps.}
    \label{fig:benign_ns_comparison}
\end{figure*}

\noindent\textbf{Benign-Nonsuspicious Apps}: Figure \ref{fig:benign_ns_comparison} illustrates the recall performance for \baseline{} and \focused{}. By averaging the priority scores of 3, 4, and 5, with a minimum threshold of 5 logs, \focused{} achieves a recall of 0.86. This indicates a high level of accuracy in identifying benign-nonsuspicious applications. In contrast, \baseline{} demonstrates a lower recall of 0.63, suggesting a decrease in detection performance.

\begin{figure*}[h!]
    \centering
    \begin{subfigure}[b]{0.5\textwidth}
        \centering
        \includegraphics[width=\textwidth]{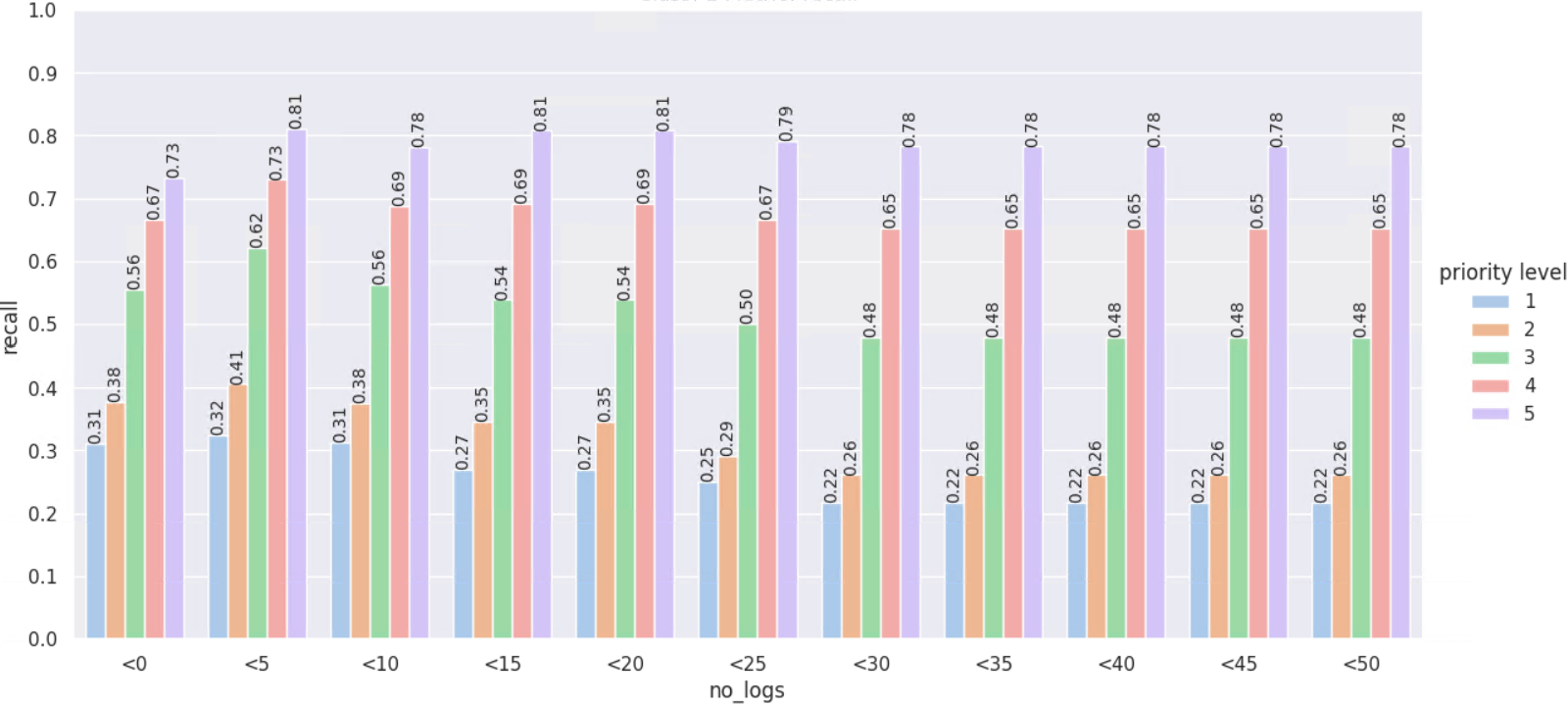}
        \caption{Recall for \baseline{} for benign-suspicious apps.}
        \label{fig:benign_s_recall_v0.7}
    \end{subfigure}%
    \begin{subfigure}[b]{0.5\textwidth}
        \centering
        \includegraphics[width=\textwidth]{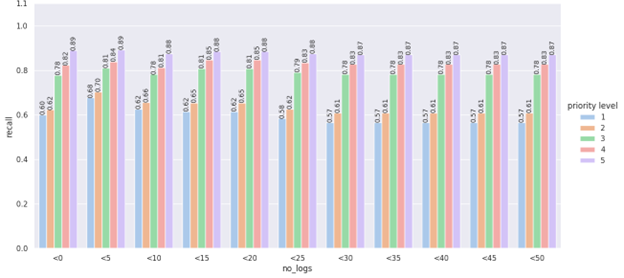}
        \caption{Recall for \focused{} for benign-suspicious apps.}
        \label{fig:benign_s_recall_v0.4}
    \end{subfigure}
    \caption{Comparison of recall for \baseline{} and \focused{} for benign-suspicious apps.}
    \label{fig:benign_s_comparison}
\end{figure*}

\noindent\textbf{Benign-Suspicious Apps}: Figure \ref{fig:benign_s_comparison} presents the recall performance for \baseline{} and \focused{}. By averaging the priority scores of 3, 4, and 5, with a minimum threshold of 5 logs, \focused{} achieves a recall of 0.84. This high recall rate indicates that \focused{} is effective in identifying benign-suspicious applications. In comparison, \baseline{} achieves a recall of 0.72, reflecting a lower detection performance.

As shown by the results, \name{} demonstrates resilience in identifying both types of applications. Although \name{} occasionally misses true positive benign applications, its high precision, as highlighted in previous sections, significantly reduces the analysts’ backlog of investigations. This high precision ensures that the majority of identified applications are indeed benign, thereby streamlining the investigative process and allowing analysts to focus on more critical tasks. The ability of \name{} to maintain such high precision while handling a diverse range of applications underscores its effectiveness and reliability in real-world scenarios.

\section{Discussion, Limitations and Future Work}
\label{sec:discussion}

In this section, we provide a comprehensive analysis of the limitations associated with \name{}, exploring the various challenges and constraints that may impact its effectiveness and usability.

\subsection{LLM inconsistencies}
\name{} leverages Large Language Models (LLMs), specifically GPT-4 Turbo, for its analytical processes. It is a well-known phenomenon that LLMs can exhibit inconsistent behaviors. While LLMs provide parameters, such as the temperature setting, to mitigate this inconsistency, even setting the temperature to 0 does not entirely eliminate variability in outputs. This issue is exacerbated when dealing with large context windows, which is often the case when working with extensive logs. Research has shown that LLMs tend to struggle as the input size increases \cite{liu2023lostmiddlelanguagemodels}.

To address these inconsistencies, \name{} performs multiple analyses per application. By running these analyses several times, \name{} achieves more robust results. To determine the priority level after multiple runs, \name{} employs a majority voting system. Our observations indicate that, in most cases, LLM outputs are consistent, with only occasional discrepancies. This approach enhances \name{}'s resilience against the inherent inconsistencies of LLMs, albeit at an increased monetary cost.

\subsection{LLM hallucination}

Hallucination is a more challenging problem than inconsistency, as it is also hard to detect. Since we are grounding LLMs with actual data, we have observed only a limited number of hallucinations, albeit an interesting one. When the LLM is provided with sign-in logs, one of the columns is the error code, and in one instance, the code is “ForkingIgnored.” The quote below shows an example excerpt from the LLM (GPT) output.

\begin{quote} Multiple instances of the error ForkingIgnored from the IP address block fd00::b918:xxxx:xxx:xx:xx. \label{quote_forking} \end{quote}

As shown in the quote, the LLM detects this ForkingIgnored error and further interprets it as a sign of malicious behavior. Below is an excerpt from asking what ForkingIgnored means:

\begin{quote} The error “ForkingIgnored” in Azure Active Directory (AAD) sign-ins typically indicates that a sign-in attempt was made from an IP address that has been marked to ignore any forked processes. This could happen in scenarios where security policies are in place to prevent sign-ins from processes that have been forked from another process, \emph{which is often a characteristic of suspicious or malicious activity}. \end{quote}

As can be seen from the quote, the LLM’s inherent knowledge of this error without grounding makes this behavior suspicious. However, this error is only a technical nuance rather than a security issue. It indicates that a slice of requests that were supposed to be sent to test have failed and have no impact in terms of security. This is more of an engineering issue than a security concern; however, the LLM exhibits significant hallucination in this case.

To alleviate this issue and similar ones, \name{} relies as much on grounding as possible. \name{} includes error codes and their explanations to avoid such behaviors from LLMs.

\subsection{LLM token size}

\name{} uses GPT-4 Turbo, which has a 128K token limit (input + output). However, we have observed that this limitation causes some issues, both in terms of fitting logs into the prompt and in reasoning as the size gets larger. With all the other grounding information that \name{} passes to the prompt, we are able to fit only 500 to 1000 rows of logs into the prompt. To be safe, we cap that limit to even lower since content can vary in size based on the application.

Another problem that we partially attribute to the token size limitation is the LLM’s reasoning capabilities. Even though the model we use is GPT-4 Turbo, which has a 128K token limit, we observe that the more we provide, the more the LLM misses out. Specifically for application analysis, one attribute that we look for in IP addresses is whether that IP has ever been used by a known, verified entity within the vicinity of the same time period. We observe that as the context gets larger, GPT has a hard time checking for these among many other things that we are asking.

To alleviate this and similar issues, \name{} adopts a multi-step review-based analysis, which involves making multiple LLM calls in sequence to reduce the number of reasoning tasks we want the LLM to perform. This approach has proven to be useful as we observe that subsequent calls usually correct mistakes caused by the first LLM call, albeit with an increased monetary cost.

By implementing this multi-step analysis, \name{} ensures more accurate and reliable results. Each step in the sequence allows the LLM to focus on a smaller subset of tasks, thereby improving its performance and reducing the likelihood of errors. This method also helps in managing the token limit more effectively, ensuring that critical information is not omitted due to space constraints.

In summary, while the 128K token limit of GPT-4 Turbo presents challenges, \name{} mitigates these by capping the number of logs and employing a multi-step analysis. This strategy enhances the LLM’s reasoning capabilities and ensures more precise and dependable outcomes, even though it may incur additional costs.

\subsection{Scalability and cost}

\name{} relies on two foundational models: \stm{} and GPT-4 Turbo. These models are integral to its functionality, with \stm{} handling data reduction and GPT-4 Turbo focusing on reasoning tasks. Both models require GPUs to operate efficiently.

\stm{} is a small language model, consisting of 700 million parameters and based on the DeBERTa-v2 architecture. It can run on V100 NVIDIA GPUs, which are well-suited for its computational needs. On the other hand, GPT-4 Turbo is a more demanding model, necessitating the use of at least NVIDIA A100 GPUs to function optimally.

\name{} leverages parallelization to efficiently process multiple applications simultaneously. This capability ensures that \name{} can scale effectively to handle a large number of applications, maintaining high performance and reliability even as the workload increases.

The costs associated with these GPUs and models are thoroughly documented in the Azure website \cite{azure_pricing}, providing transparency and detailed information for budgeting and planning purposes.

\subsection{Limitations and future improvements}

\name{} was initially designed to analyze applications. However, its modular design allows for expansion to analyze various entities, including but not limited to devices, users, or any other service principals. While \name{}'s results are promising, we acknowledge that the number of applications used in our experiments is limited and needs to be expanded. For further improvements, we will focus on three key areas:

\textbf{Reasoning with Graphs}: The initial design of \name{} relies on typical Retrieval Augmented Generation (RAG) techniques, which bring context to the prompt. To enhance \name{}'s reasoning capabilities, we plan to explore the use of GraphRAG \cite{edge2024localglobalgraphrag}, which utilizes a knowledge graph derived from the underlying data to help answer questions. Various aspects will be considered, such as logs, threat intelligence (e.g., TA profiles), and the overall architecture of \name{}. This includes examining how the outputs of different components of \name{} can be integrated into GraphRAG for improved reasoning.

\textbf{Reasoning with Multiple Agents}: The initial design of \name{} includes multiple steps that pulls relevant data, reduce the size, reasons on it and review. LLM-based agents have proven useful in correcting mistakes and achieving better outcomes. We will explore the use of multiple agents to reduce false positives and enhance the performance of \name{}.

\textbf{Data Reduction}: One of the main challenges of working with logs and LLMs is the large volume of data. \name{} currently relies on subsampling and anomaly detection to focus on the data that matters. We will work to improve and optimize these aspects, which is also crucial for the system’s scalability.
\section{Conclusion}
\label{sec:conclusion}

This paper introduces \name{}, detailing its design, experiments, limitations, and future work. \name{} achieves high precision and recall in detecting malicious applications by leveraging language models. In our experiments, \name{} achieved an impressive F-1 score of 0.90 in detecting malicious applications, significantly reducing the case load for security analysts by 87\% in a real-world use case.

For future versions of \name{}, we plan to focus on enhancing its reasoning capabilities and scalability. This will involve improving its ability to understand and interpret complex patterns in data, as well as ensuring it can handle larger datasets and more diverse types of malicious applications. By addressing these areas, we aim to make \name{} an even more powerful tool for cybersecurity professionals.

\small
\bibliographystyle{plain}
\bibliography{paper.bib}

\clearpage
\begin{appendices}

\section{Example Threat Actor Profile}
\label{sec:app:profile}

\begin{multicols}{2}
\begin{tcolorbox}[colframe=black, width=\textwidth, arc=0mm, outer arc=0mm]
\begin{lstlisting}[language=yaml, caption={Example YAML file for Threat Actor Profile}]
profile: |
	- Initial Access:
            - Signin to sensitive resources.
        - Execution:
            - Calls made to Graph API to perform unusual actions.
        - Persistence:
            - Creation of user or application in tenant YYY.
        - Reconnaissance:
            - Microsoft Graph activity
            - Microsoft Data Explorer (Kusto) activity
        - Privilege Escalation:
        - Defense Evasion:
            - Use of proxy services.
        - Credential Access:
            - Sign-in to Azure Key Vault resource.
            - List operation followed by Get operations in a Key Vault.
        - Lateral Movement:
            - First access to an unusual resource. 
        - Data Collection:
            - Data access from storage resources.
\end{lstlisting}
\end{tcolorbox}
\end{multicols}

\section{Example \name{} Output}
\label{sec:app:output}

\begin{multicols}{2}
\begin{tcolorbox}[title=High level behavior of the application, colframe=black, width=\textwidth, arc=0mm, outer arc=0mm]
\normalsize
The application, identified as "ContosoApp," primarily interacts with Microsoft Graph, which is a gateway to data and intelligence in Microsoft 365. It performs various operations such as retrieving transitive members of groups, checking member groups, and accessing service principal information. The application has permissions to read directory data, get secrets from Key Vault, and read all SharePoint sites. The application's activities are spread across multiple IP addresses, and it has been observed accessing sensitive resources like Microsoft Graph and Azure Key Vault.
\end{tcolorbox}
\end{multicols}

\clearpage

\begin{multicols}{2}
\begin{tcolorbox}[breakable, title=Suspicious activities and entities involved, colframe=black, width=\textwidth]
\normalsize

Based on the provided logs and the threat actor profile, the following suspicious activities have been identified:

\textbf{Initial Access}
\begin{itemize}
    \item \textbf{Evidence:} There are sign-in attempts from IP address \texttt{XX.XX.XX.X} which is not listed in the benign IP addresses and is associated with a proxy service (\texttt{XXX\_PROXY}). This IP has accessed Microsoft Graph and attempted to create users with display names "Bob Smith" and "Bob Smith 2" and userPrincipalNames \texttt{bob.smith@yourdomain.com} and \texttt{bob.smith2@yourdomain.com}, respectively. This activity is indicative of the threat actor using stolen credentials to infiltrate the network.
    \item \textbf{Entities:} IP \texttt{XX.XX.XX.X}, UserPrincipalNames \texttt{bob.smith@yourdomain.com}, \texttt{bob.smith2@yourdomain.com}
    \item \textbf{Dates:} 2024-02-23
\end{itemize}

\textbf{Execution}
\begin{itemize}
    \item \textbf{Evidence:} The same suspicious IP \texttt{XX.XX.XX.X} has been observed using the Microsoft Graph API to interact with service principals and users, which could be an attempt to execute commands or queries within the Microsoft environment.
    \item \textbf{Entities:} IP \texttt{XX.XX.XX.X}
    \item \textbf{Dates:} 2024-02-23
\end{itemize}

\textbf{Persistence}
\begin{itemize}
    \item \textbf{Evidence:} The access to Azure Key Vault by IP \texttt{XX.XX.XX.X} to retrieve the secret named "ContosoAppOfficeKey" suggests an attempt to maintain access to sensitive resources.
    \item \textbf{Entities:} IP \texttt{XX.XX.XX.X}, Key Vault secret "ContosoAppOfficeKey"
    \item \textbf{Dates:} 2024-02-23
\end{itemize}

\textbf{Privilege Escalation}
\begin{itemize}
    \item \textbf{Evidence:} The creation of new users in the Microsoft Graph by IP \texttt{XX.XX.XX.X} could be an attempt to escalate privileges within the organization.
    \item \textbf{Entities:} IP \texttt{XX.XX.XX.X}, New users "Bob Smith" and "Bob Smith 2"
    \item \textbf{Dates:} 2024-02-23
\end{itemize}

\textbf{Defense Evasion}
\begin{itemize}
    \item \textbf{Evidence:} The use of a proxy service (\texttt{XXX\_PROXY}) by IP \texttt{XX.XX.XX.X} indicates an attempt to evade defenses by masking the true origin of the attack.
    \item \textbf{Entities:} IP \texttt{XX.XX.XX.X}
    \item \textbf{Dates:} 2024-02-23
\end{itemize}

\textbf{Credential Access}
\begin{itemize}
    \item \textbf{Evidence:} The retrieval of the secret "ContosoAppOfficeKey" from Azure Key Vault by IP \texttt{XX.XX.XX.X} suggests that the threat actor has gained access to sensitive credentials.
    \item \textbf{Entities:} IP \texttt{XX.XX.XX.X}, Key Vault secret "ContosoAppOfficeKey"
    \item \textbf{Dates:} 2024-02-23
\end{itemize}

\textbf{Data Collection}
\begin{itemize}
    \item \textbf{Evidence:} The access to Microsoft Graph and the attempt to retrieve transitive members of groups could be indicative of data collection activities by the threat actor.
    \item \textbf{Entities:} IP \texttt{XX.XX.XX.X}
    \item \textbf{Dates:} 2024-02-23
\end{itemize}
\end{tcolorbox}
\end{multicols}

\clearpage

\begin{multicols}{2}
\begin{tcolorbox}[title=Triage priority level of the application, colframe=black, width=\textwidth, arc=0mm, outer arc=0mm]
\normalsize
\begin{enumerate}
    \item 'Initial Access' is observed. [True]
    \item 'Execution' is observed. [True]
    \item 'Persistence' is observed. [True]
    \item 'Reconnaissance' is observed. [False]
    \item 'Privilege Escalation' is observed. [True]
    \item 'Defense Evasion' is observed. [True]
    \item 'Credential Access' is observed. [True]
    \item 'Lateral Movement' is observed. [False]
    \item 'Data Collection' is observed. [True]
    \item Application has extensive permissions, and can access sensitive resources. [True]
    \item More than one stages of Kill Chain is observed. [True]
    \item All access attempts are unsuccessful and resulted with errors. [False]
    \item Application did not access sensitive resources. [False]
    \item All of the suspicious IP addresses are benign. [False]
    \item All of the suspicious IP addresses were linked to resources that are not considered high value (check IP Details table Resource Accessed column). This makes the IP address less suspicious. [False]
\end{enumerate}
\end{tcolorbox}
\end{multicols}

\end{appendices}

\end{document}